\documentclass[hyper]{JHEP3}

\input{epsf}
\usepackage{epsfig}
\usepackage{amssymb}
\usepackage{amsfonts}
\usepackage{amsbsy}
\usepackage[all]{xy}
\usepackage{amsmath}

\usepackage{amssymb,amscd}
\usepackage{mathrsfs}
\usepackage{amsmath,amsthm}

\usepackage{slashed}

\makeatletter
\DeclareFontFamily{OMX}{MnSymbolE}{}
\DeclareSymbolFont{MnLargeSymbols}{OMX}{MnSymbolE}{m}{n}
\SetSymbolFont{MnLargeSymbols}{bold}{OMX}{MnSymbolE}{b}{n}
\DeclareFontShape{OMX}{MnSymbolE}{m}{n}{
    <-6>  MnSymbolE5
   <6-7>  MnSymbolE6
   <7-8>  MnSymbolE7
   <8-9>  MnSymbolE8
   <9-10> MnSymbolE9
  <10-12> MnSymbolE10
  <12->   MnSymbolE12
}{}
\DeclareFontShape{OMX}{MnSymbolE}{b}{n}{
    <-6>  MnSymbolE-Bold5
   <6-7>  MnSymbolE-Bold6
   <7-8>  MnSymbolE-Bold7
   <8-9>  MnSymbolE-Bold8
   <9-10> MnSymbolE-Bold9
  <10-12> MnSymbolE-Bold10
  <12->   MnSymbolE-Bold12
}{}

\let\llangle\@undefined
\let\rrangle\@undefined
\DeclareMathDelimiter{\llangle}{\mathopen}%
                     {MnLargeSymbols}{'164}{MnLargeSymbols}{'164}
\DeclareMathDelimiter{\rrangle}{\mathclose}%
                     {MnLargeSymbols}{'171}{MnLargeSymbols}{'171}
\makeatother

\def\be{ \begin{equation} }
\def\ee{ \end{equation}}

\def\dim{{\rm dim}}
\def\exp{{\rm exp}}

\def\Hom{{\rm Hom}}
\def\I{{\rm i}}

\renewcommand{\Im}{{\rm Im }}
\def\ker{{\rm ker}}
\def\Lie{{\rm Lie}}

\def\mod{{\rm mod}}

\def\Tr{{\rm Tr}}

\def\vol{{\rm vol\,}}

\def\half{\frac{1}{2}}
\def\p{\partial}

\def\one{{\hbox{ 1\kern-.8mm l}}}

\def\CA{{\cal A}}
\def\CB{{\cal B}}

\def\CE {{\cal E}}

\def\CG {{\cal G}}
\def\CH {{\cal H}}

\def\CK {{\cal K}}
\def\CL {{\cal L}}
\def\CM {{\cal M}}

\def\CR {{\cal R}}

\def\CX {{\cal X}}

\def\CE {{\cal E}}
\def\CG {{\cal G}}
\def\CH {{\cal H}}

\def\CB {{\cal B}}
\def\CQ {{\cal Q}}
\def\CS {{\cal S}}

\def\CU {{\cal U}}
\def\CX {{\cal X}}
\def\CY {{\cal Y}}

\def\IC{\mathbb{C}}

\def\IJ{\mathbb{J}}

\def\IP{\mathbb{P}}

\def\IR{{\mathbb{R}}}

\def\IZ{{\mathbb{Z}}}

\def\fg{\mathfrak{g}}

\def\fl{\mathfrak{l}}

\def\fs{\mathfrak{s}}
\def\ft{\mathfrak{t}}

\def\fs{\mathfrak{s}}
\def\ft{\mathfrak{t}}
\def\fu{\mathfrak{u}}

\def\fP{\mathfrak{P}}
\def\fQ{\mathfrak{Q}}

\def\rmG{{\mathrm{G}}}

\def\rmk#1{\bigskip\noindent{\bf Remarks} }

\newcommand\fro{{\overline{\underline{\Omega}}}}
\def\hk{hyperk\"ahler}
\def\froM{ \overline{\underline{\CM}} }
\def\fMM{\overline{\underline{\mathcal{M}}}}
\def\Dsl{\slashed{D}}

\title{A Note On The Semiclassical Formulation Of  BPS States In Four-Dimensional N=2 Theories }

\author{T. Daniel Brennan  and  Gregory W.~Moore  \\
  NHETC and
Department of Physics and Astronomy, Rutgers University \\
126 Frelinghuysen Rd., Piscataway NJ 08855, USA\\
{\tt tdanielbrennan@physics.rutgers.edu,gmoore@physics.rutgers.edu} }

\abstract{Vector spaces of (framed) BPS states of Lagrangian four-dimensional N=2 field theories can be defined 
in semiclassical chambers in terms of the $L^2$-cohomology of Dirac-like operators on monopole moduli 
spaces. This was spelled out in   \cite{Moore:2015qyu,Moore:2015szp} for theories with only 
vectormultiplets,   taking into account only a subset of the possible 
half-supersymmetric 't Hooft-Wilson line defects. This note completes the discussion by describing 
the modifications needed when including matter hypermultiplets together with arbitrary 't Hooft-Wilson line defects. Two applications of this extended discussion are given.  \today  }

\begin{document}

\section{Introduction And Conclusion}\label{sec:Intro}

This note summarizes part of a talk delivered by one of us at the 
Nambu Memorial Symposium at the University of Chicago, March 2016. 
Professor Nambu's profound contributions to the theory of spontaneous symmetry 
breaking and to the use of nonabelian gauge theory in particle physics 
firmly establishes him as one of the great physicists of the twentieth 
century. This note is about the theory of magnetic monopoles. As far 
as we know, Professor Nambu never wrote a paper about magnetic monopoles, 
but given that these are a beautiful and important aspect of 
spontaneously broken nonabelian gauge theories the topic seems to 
us to be most apt as a contribution to this memorial volume for 
Professor Nambu. 
 
BPS states \cite{Seiberg:1994rs,Witten:1978mh} (and their framed analogues \cite{Gaiotto:2010be}) in four-dimensional N=2 supersymmetric 
field theories can be defined in the semiclassical limit in terms of $L^2$ cohomology of Dirac-like 
operators on moduli spaces of (singular) monopoles, as is well-known from a rather extensive 
previous literature. Some finishing touches of this formulation 
were recently worked out in \cite{Moore:2015qyu} but only for pure gauge theories, and only 
for  framed BPS states in the presence of a subset of the possible 't Hooft-Wilson line defects.  
This note extends the finishing touches of \cite{Moore:2015qyu} to include theories with arbitrary 
hypermultiplet representations together with 
arbitrary 't Hooft-Wilson line defects. The main modification is that the 
Dirac operator must be coupled to a hyperholomorphic vector bundle.  The relevant hyperholomorphic 
bundle is described in section \ref{sec:HHBundles} and is derived from the universal bundle of Atiyah and Singer \cite{Atiyah:1984tf}. The proof of the main claim 
follows easily by constructing the relevant N=4 supersymmetric quantum mechanics on  
monopole moduli space, following 
\cite{Gauntlett:1993sh,Gauntlett:1995fu,Gauntlett:1999vc,Gauntlett:2000ks,Moore:2015szp,Sethi:1995zm,
Stern:2000ie,Weinberg:2006rq}. Some details are in appendix \ref{sec:SQM}.

There are two applications of this work. 
First, given the truth of the no-exotics conjecture of \cite{Gaiotto:2010be}  (partially proven in 
\cite{Chuang:2013wt,CordovaDumitrescu,DelZotto:2014bga}), the arguments here complete the proof of the 
Generalized Sen Conjecture described in section \ref{sec:GenSen} below and in 
section 4.1 of \cite{Moore:2015qyu}. Second, as in \cite{Moore:2015szp}, when combined 
with explicit computations from \cite{Gaiotto:2010be} we obtain a wealth of predictions for the $L^2$ 
kernels of Dirac operators on (singular) monopole moduli spaces. The novel point in this 
note is that these predictions are extended to Dirac operators coupled to 
certain hyperholomorphic bundles described below.  One example is worked out 
in detail in section \ref{sec:Explicit}.

This paper builds on and extends the letter  \cite{Moore:2015qyu}. We will 
endeavor to use the same notation as in that letter and in the interest of brevity 
we will not always fully define notation - so we  
we will assume the reader has some familiarity with \cite{Moore:2015qyu}.    
Full computations and complete details 
can be found in the forthcoming PhD thesis of the first author. Further background 
material and extensive references 
to the large  literature on the semiclassical formulation of BPS states can be 
found in \cite{Moore:2015szp,Weinberg:2006rq}.

\section*{Acknowledgements}

We thank Anindya Dey, Andy Royston, David Tong,  Dieter Van den Bleeken, Edward Witten,
and Kenny Wong and  for correspondence and discussions on related material. Special 
thanks go to Andy Royston for detailed comments on a preliminary version of the draft. 
This work is supported by the U.S. Department of Energy under grant DOE-SC0010008 to 
Rutgers University. GM thanks the organizers of the Nambu Memorial Conference at 
the University of Chicago for the invitation to speak. 
The work of G.M. is also supported by the IBM Einstein Fellowship of the Institute for Advanced Study.

\section{Statement And Solution Of The Problem}\label{sec:Statement}

We wish to describe BPS states in a semiclassical limit, allowing for the presence of 
arbitrary 't Hooft-Wilson line defects. We thus confine attention to Lagrangian 
 d=4 N=2 theories. These are described by the following data: 

\begin{enumerate}

\item \emph{Gauge group and couplings}: A semisimple compact Lie group $G$ together with a 
complex gauge coupling $\tau$ for each simple factor of $G$. 

\item \emph{Matter hypermultiplets}: A quaternionic representation $\CR$ of $G$ 
compatible with a positive inner product on $\CR$. (Thus, all the complex structures 
are orthogonal transformations of $\CR$.)   

\item \emph{Mass parameters}: The \emph{flavor group} $G_f$ is defined to be the 
commutant of $G$ in the orthogonal group $O(\CR)$ preserving the inner product on 
$\CR$. The mass parameters $m$ are valued in 
$\fg_f \otimes \IC$ where $\fg_f$ is the Lie algebra of $G_f$. Then N=2 supersymmetry 
requires   $[m, m^\dagger]=0$. Hence we can assume that $m \in \ft_f\otimes \IC$ 
is in a Cartan subalgebra of $G_f$. We will further assume that it is a regular element 
so that the flavor symmetry group is broken to a maximal torus $\cong U(1)^{N_f}$ by the masses.

\end{enumerate}

A quick and elegant way to understand that this is the appropriate 
way to formulate mass parameters is to use the viewpoint \cite{Argyres:1996eh,Seiberg:1994aj}
that $m$ 
are the vev's of adjoint scalars of vectormultiplets when the flavor group $G_f$ 
is weakly gauged (i.e. the flavor gauge coupling is taken to zero).  
The vacuum condition for these scalars is simply 
$[m, m^\dagger]=0$.

Next we need the data defining half-supersymmetric 't Hooft-Wilson line defects. 
These are determined by the data: 

\begin{enumerate}

\item \emph{Unbroken supersymmetry}:  A choice of  phase $\zeta\in U(1)$ (or rather a lift
of $\zeta$  to the universal cover of $U(1)$) specifying which
 four supersymmetries of the half-supersymmetric defect remain unbroken. 
 For  details see \cite{Gaiotto:2010be,Kapustin:2006hi}.

\item \emph{'t Hooft-Wilson charges}: 
An equivalence class of a pair $[P,Q]$ where $P$ is a cocharacter of $G$ and 
$Q$ is a  weight of the centralizer $Z(P) \subset G$. The square brackets 
indicate the equivalence class under the diagonal action of the Weyl group of $G$. 
Using this data we can define defect boundary conditions on the field in the path integral
\cite{Kapustin:2005py}.

\end{enumerate}

We denote the line defect determined by the above data by $L_\zeta[P,Q]$. 

Finally, we need infrared data. These consist of

\begin{enumerate}

\item \emph{A Coulomb branch vacuum}:
The Coulomb branch is    $ \CB := \ft\otimes\IC/W$ where $\ft$ is a Cartan subalgebra of $\fg=\Lie(G)$. 
A typical point is traditionally denoted $u\in \CB$. 
The ``semiclassical region'' is a set of regions where $u\to \infty$ on the Coulomb branch. 
The precise definition can be found in section 4.6 of \cite{Moore:2015szp} but the basic 
idea is very simple: One takes the bare coupling constant to zero and hence $\Im(\tau)\to \infty$ 
for each simple factor of $G$. A point $u$ on the Coulomb branch vacuum determines a 
vacuum expectation value $X_\infty\in \fg $ of a Higgs field $X$ up to conjugation. 
(The field $X$ is defined in  equation \eqref{eq:XY-Field-Def}.) We will assume $X_\infty$ is a regular element 
of $\fg$ and hence determines a Cartan subalgebra $\ft$, a Cartan subgroup $T\subset G$, 
and a set of positive roots.  

\item \emph{An infrared charge}: The mass parameters and vacuum expectation value $X_\infty$ 
break the $G_f \times G$ symmetry to an abelian group. Taking into account dual magnetic 
symmetries, the symmetry group of the IR theory is a torus $ \tilde T$ that fits in an exact sequence
\be
1 \rightarrow T_{em} \rightarrow \tilde T  \rightarrow T_f \rightarrow 1.
\ee
Here $T_f$ is the Cartan subgroup of the flavor group while $T_{em}$ is the group of electric 
\underline{and} magnetic gauge transformations. 
The IR charges, $\gamma$, are in the character lattice $\Gamma$ of $\tilde T$ and hence also fit in a sequence: 
\be\label{eq:GammaSeq}
0 \rightarrow \Gamma_f \rightarrow \Gamma \rightarrow \Gamma_{em} \rightarrow 0 .
\ee
Here the lattice of flavor charges $\Gamma_f$ is just the character lattice 
of the unbroken flavor symmetry $T_f$ while $\Gamma_{em}$ is the symplectic lattice of 
electric and magnetic gauge charges. The above sequences split, but in general not naturally 
since one can add a gauge current to a flavor current. 

\end{enumerate}

Given the above data one can formulate the general problem: \emph{Define and 
compute the vector space of framed BPS states for the theory in question with the 
specified IR data.}  This is an extremely difficult problem and has been the subject 
of much research. However, when   $u$ is in a weak-coupling region the problem is much more 
manageable, although it still requires a little attention to give a precise statement. The 
full solution to the semiclassical version of this problem is the subject of this note. 

When the problem is restricted to theories consisting only of vectormultiplets, 
together with a subset of the possible 't Hooft-Wilson line defects (this subset 
includes $L_\zeta[P,0]$ for all $P$)  the solution was 
explained in \cite{Moore:2015qyu,Moore:2015szp}. We summarize the answer very briefly. 
To begin, in the semiclassical regime there is a distinguished family of duality frames, all 
related by the Witten effect.  
There is a   canonical splitting of \eqref{eq:GammaSeq} 
(up to the Witten effect) which allows us to decompose a charge $\gamma\in \Gamma$ as 
\be
\gamma = \gamma_f \oplus \gamma_m \oplus \gamma_e \in \Lambda_{wt,f} \oplus \Lambda_{mw} 
\oplus \Lambda_{wt}.
\ee
We will denote $\gamma_{e+f}:= \gamma_f \oplus \gamma_e$ below. The Witten effect arises from 
monodromy defined by a map $\Lambda_{mw}  \rightarrow \Lambda_{wt,f}\oplus  \Lambda_{wt}$, 
but the magnetic charge $\gamma_m$ is invariant in the semiclassical regime.

In the semiclassical region it is useful to define  a pair of ``real'' adjoint vevs \cite{Moore:2015szp}
\be\label{eq:XYdef}
\begin{split} 
\CX & := {\rm Im}(\zeta^{-1} a(u)) \in \ft \\
\CY & := {\rm Im}(\zeta^{-1} a_D(u)) \in \ft \\
\end{split}
\ee
where $a(u)$ and $a_D(u)$ are the periods relative to the canonical weak coupling duality frame
and $\CX = X_\infty+\cdots $ to leading order in the weak coupling expansion. 
When there is no line defect we apply the same formula with  $\zeta = - Z^{cl}/\vert Z^{cl} \vert$ in the 
weak coupling limit (see equation \eqref{eq:Zdef} below).

Using the vev $\CX$ and the magnetic charge $\gamma_m$ we can 
define a moduli space of (possibly singular) magnetic monopoles.
\footnote{We follow the notation of \cite{Moore:2014jfa,Moore:2014gua,Moore:2015qyu,Moore:2015szp} for the monopole moduli spaces. 
Thus $\froM([P], \gamma_m; \CX)$ is the moduli space of singular monopoles with  't Hooft 
charge $[P]$, magnetic charge $\gamma_m$ and Higgs vev $\CX$ at infinity. When $P=0$ 
we simply write $\CM(\gamma_m ; \CX)$. Often we simply write $\froM$ and $\CM$ if the 
arguments are understood.} 
Then, the semiclassical dynamics of BPS states with magnetic charge $\gamma_m$ are 
described by  collective coordinates on the moduli space. These collective coordinates are governed by an N=4 supersymmetric quantum mechanics,  and one 
  of the supersymmetry operators is the  Dirac operator 
\be\label{eq:DiracOperator}
\Dsl^\CY = \Dsl + \slashed{\rmG}(\CY)
\ee
where $\Dsl$ is the ordinary Dirac operator acting on Dirac spinors on $\froM$ or $\CM$ 
and $\slashed{\rmG}(\CY)$ is Clifford multiplication by the hyperholomorphic vector field associated with global gauge transformations 
by $\CY \in \ft$. 
\footnote{See \cite{Moore:2015qyu,Moore:2015szp} for more details about ${\rmG}(\CY)$. Briefly, 
there is a Lie algebra homomorphism from $\ft$ to the hyperholomorphic vectorfields on moduli 
space $H \mapsto \text{G}(H)$ implementing the action of an infinitesimal global gauge transformation by $H$.
The relevant gauge transformation is defined under equation \eqref{eq:phimn} below. }
One then defines the space of all framed BPS states with 
fixed \underline{magnetic} charge $\gamma_m$,
in the presence of the line defect $L_\zeta[P,0]$,  to be the $L^2$ kernel, denoted here 
by $\CK$,  of the operator  $\Dsl^{\CY}$ on $\froM([P], \gamma_m;\CX) $:
\be
\CK := {\rm ker}_{L^2} \Dsl^{\CY}. 
\ee
The space $\CK$ is a representation of a  group isomorphic to 
\be
 T \times SO(3)_{\rm rotation} \times SU(2)_R, 
\ee
where $T$ is the maximal torus of $G$ determined by the commutant of the regular vev 
$\CX$,  $SO(3)_{\rm rotation}$ is the group of rotations around some point in spatial $\IR^3$,  and  
$SU(2)_R$ is the commutant of the symplectic holonomy of the \hk\ metric. The group $SU(2)_R$ 
has a lift to the spin bundle and preserves  $\CK$. The group 
$SO(3)_{\rm rotation}$ induces a group of isometries of the \hk\ metric and again 
preserves  $\CK$. Finally, global gauge transformations by $T$ 
are hyperholomorphic and commute with $\Dsl^\CY$.  
The isotypical subspaces of  $\CK$, when decomposed as a $T$-representation,  
are identified with the subspaces of framed BPS states of definite electric charge.  They therefore are in 
representations of $SO(3)_{\rm rotation} \times SU(2)_R$.

Now, the modification of the above  answer in the case where we include general 't Hooft-Wilson lines 
(including the possibility $P=0$ and $Q\not=0$, i.e. general Wilson lines) as well as   general matter hypermultiplets is very simple. 
One defines an Hermitian hyperholomorphic vector bundle $\CE_{\rm line}$ associated with the line defects  together with a  hyperholomorphic bundle $\CE_{\rm matter}$ associated with the quaternionic representation 
$\CR$. Then we simply use the same Dirac operator as before coupled to 
\be\label{eq:TotalCE}
\CE = 
\CE_{\rm line} \otimes {\rm Spin}(\CE_{\rm matter}) 
\ee
where  ${\rm Spin}(\CE_{\rm matter}) $ is a vector bundle associated to the 
spin bundle of $\CE_{\rm matter}$. The bundle $\CE_{\rm matter}$ represents hypermultiplet 
fermion degrees of freedom in the supersymmetric quantum mechanics of appendix 
\ref{sec:SQM} and, upon quantization of the Clifford algebra based on $\CE_{\rm matter}$,
we obtain states in the spin representation. 
The bundle $\CE$ inherits a hyperholomorphic connection from the ones on  
  $\CE_{\rm line}$  and $\CE_{\rm matter}$. These bundles and connections are defined in section \ref{sec:HHBundles} below. 
To define (framed) BPS states of definite magnetic charge we
 take the $L^2$ kernel of this operator on $\froM$ if $P\not=0$ and on $\CM$ if 
 $P=0$ but $Q\not=0$. If there are no line defects we must 
take the $L^2$ kernel on the strongly-centered moduli space (a factor of the univeral cover) 
and impose an equivariance 
condition under the action of the Deck group on the universal cover. These complications are explained at length 
in \cite{Moore:2015qyu,Moore:2015szp} and no new issues arise in the more general situation 
we consider here. 

Once again, the  torus   $  T_f \times T$ of the unbroken flavor and gauge symmetry acts on the 
bundle $\CE$ and commutes with the Dirac operator. Therefore 
the $L^2$-kernel is a representation of 
\be
T_f \times T  \times SO(3)_{\rm rotation} \times SU(2)_{R} .
\ee
 The flavor and 
electric charges are determined by the character of $T_f \times T$ acting on the kernel. The desired space 
of BPS states is the isotypical subspace: 
\be
\overline{\underline{\CH}}( L_\zeta[P,Q]; \gamma;u):= \ker_{L^2}^{\gamma_{e+f}}(\Dsl^\CY)
\ee
in the framed case, with a similar equation for the vanilla case (i.e. without line defects). 

\section{Construction Of The Hyperholomorphic Bundles}\label{sec:HHBundles}

\subsection{Hyperholomorphic Bundles Associated To Line Defects}\label{subsec:HHBundles:LD}

We suppose a line defect $L_\zeta[P,Q]$ has been inserted at a point $\vec x\in \IR^3$. 
Let $\CQ$ denote the universal principal $Z(P)$-bundle of appendix \ref{app:Universal} 
over $\IR^3 \times \CA^*/\CG$. 
We can pull back 
the bundle using $\iota_{\vec x}: \{\vec x \} \times \froM \hookrightarrow \IR^3 \times \CA^*/\CG$ 
to obtain a principal $Z(P)$ bundle over $\froM$. The Wilson line data $Q$ defines a representation 
$R(Q)$ of $Z(P)$ and we then form the associated vector bundle for this representation. The 
bundle $\CE_{\rm line}(\vec x;Q)$ is defined to be this associated bundle. The universal connection 
pulls back to a hyperholomorphic connection on $\CE_{\rm line}(\vec x;Q)$. The simplest proof 
that it is hyperholomorphic, for a physicist, follows from the existence of the N=4 supersymmetric 
quantum mechanics of appendix \ref{sec:SQM}.  In the case when $P=0$,  $\CE_{\rm line}(\vec x;Q)$ is a bundle 
over $\CM$.

 One can of course consider the insertion of multiple line defects.  
 If there are several  defects $L_\zeta[P_j,Q_j]$ inserted at  points $\vec x_j$,
  all preserving the same supersymmetry,  then we simply 
have a bundle   associated to each point and in the definition of framed BPS states 
we take the tensor product over all points: 
\be
\CE_{\rm line} := \otimes_{j} \CE_{\rm line}(\vec x_j;Q_j) .
\ee

\subsection{Hyperholomorphic Bundles Associated To Hypermultiplet Matter}\label{subsec:HHBundles:HM}

When including hypermultiplet matter in a quaternionic representation $\CR$ with mass 
parameters $m$ we define a hyperholomorphic bundle $\CE_{\rm matter}$ over 
monopole moduli space. We do this by considering the trivial Hilbert bundle 
$\CA^* \times \CH$ where $\CH$ is a Hilbert space of $L^2$-sections of a 
spin-bundle over $\IR^3$ (coupled to a vector bundle). The bundle is $\CG$-equivariant 
and descends to a bundle with connection on $\CA^*/\CG$ where the connection 
is the ``universal connection'' described in equation \eqref{eq:Hproj} below.  We pull this 
back to the monopole moduli subspace and 
project to the kernel of a certain Dirac operator $L$ (not to be confused 
with $\Dsl^{\CY}$).  Using a Bochner-type argument 
the kernel of $L$ does not jump as we vary the parameters over $\fMM$ and 
the resulting vector bundle with its projected connection is the hyperholomorphic 
bundle $\CE_{\rm matter}$. We now expand on the above with a few more details.

The derivation of the collective coordinates for the hypermultiplet fermion zeromodes 
(see appendix \ref{sec:SQM} below) involves finding $L^2$ solutions of a Dirac equation on 
$\IR^3$ for spinors in $\CS \otimes E_{\CR}$ where $E_{\CR} \to \IR^3$ is the bundle 
associated to the principal $G$-bundle $\fP\to \IR^3$ via the representation $\CR$, 
and $\CS$ is the spinor bundle on $\IR^3 \times \IR$ restricted to $\IR^3$. 
The Dirac operator has the form 
\be
\I \Gamma^a \hat D_a  + \I \Gamma^4 m_x
\ee
where the index $a$ runs from $1$ to $4$, $\hat D_a$ is the spinor covariant derivative coupled 
to the gauge field $\hat A$ of equation \eqref{eq:Ahat}, and we use the phase $\zeta$ to define ``real'' mass 
parameters 
\be
\zeta^{-1} m = m_y + \I m_x 
\ee
where $m_y, m_x \in \ft_f$.  Here $\Gamma^a$ are four Hermitian Dirac representation matrices and we can 
choose a representation of the form 
\be
\Gamma^a = \begin{pmatrix} 0 & \tau^a \\  \bar \tau^a & 0 \\ \end{pmatrix}
\ee
with $\tau^a = ( \vec \sigma, - \I \textbf{1})$, $\bar\tau^a = ( \vec \sigma, \I \textbf{1})$, so that 
\be
\I \Gamma^a \hat D_a + \I  \Gamma^4 m_x = \begin{pmatrix} 0 & L^\dagger \\ 
L & 0 \\  \end{pmatrix}
\ee
Using the Bogomolnyi equations one finds that $L L^\dagger$ is a sum of two positive 
semidefinite operators and thus will not have an $L^2$ kernel so we are only interested in 
the kernel of $L = \I \bar \tau^a \hat D_a - m_x$. This operator acts on the Hilbert space 
$\CH$ of $L^2$ sections of  $ S \otimes E_{\CR} \to \IR^3$ where $S$ is the spinor bundle 
of $\IR^3$. Now, in the definition of the collective coordinates, 
the dimension of $\ker L$ as a \underline{complex} vector space is the 
same as the dimension of the fibers of $\CE_{\rm matter}$ as a \underline{real} vector 
space.

The rank of $\CE_{\rm matter}$ follows from a  computation analogous to \cite{Callias:1977kg,Moore:2014jfa,Weinberg:1979ma}:
\begin{align}\begin{split}
\text{rnk}_{\mathbb{R}}[\CE_{\rm matter}]= \half \sum_{\mu\in \Delta_{\CR}}n_{\CR}(\mu)\left\{\langle \mu,\gamma_m\rangle\text{ sign}(\langle\mu,m_x  \oplus \CX\rangle )+\sum_{j}  \vert 
\langle \mu,P_j\rangle \vert \right\}
\end{split}\end{align}
where we consider the representation $\CR$ to be a representation of $G_f\times G$ and 
we sum over the weights in $\ft_{\rm flavor}^\vee \oplus \ft_{\rm gauge}^\vee$ of $\CR$. Here $n_{\CR}(\mu)$ 
is the (complex) dimension of the $\mu$-weight space. We have also 
included the possibility of having more than one line defect with 't Hooft charge $P_j$ 
located at points $\vec x_j$.  

In the 
important special case where $\CR \cong \rho \oplus \rho^*$ is a sum of two irreducible 
complex representations of $G$
the flavor group $G_f \cong U(1)$ so $\ft_f \cong \I \IR$ and the mass parameter  
becomes a complex number (with $m_x, m_y$ both pure imaginary) and we have  
\begin{align}
\begin{split}
\text{rnk}_{\mathbb{R}}[\CE_{\rm matter}]= \sum_{\mu\in \Delta_{\rho}}n_{\rho}(\mu)\left\{\langle \mu,\gamma_m\rangle\text{ sign}(\langle\mu,\CX \rangle- \I m_x )+\sum_{j} \vert  \langle \mu,P_j \rangle\vert\right\}
\end{split}\end{align}\label{eq:ComplexHM-DIM}
where we now just sum over the weights of $\rho$ as a complex $G$-representation.  

As a consistency check, consider the case of an $SU(2)$ gauge theory with $N_f$ hypermultiplets in the fundamental representation with charge $\gamma_m = k H_{\alpha}$ 
where $k$ is a positive integer.
\footnote{Here and below we denote the positive root of $\fs\fu(2)$ by $\alpha$ and 
the corresponding coroot by $H_{\alpha}$.}
Let us take $\CX= v H_{\alpha} $, 
with $v>0$  and  $-\I m_x = m_r \in \IR$.  
Then 
\begin{align}\begin{split}
\text{rnk}_{\mathbb{R}}[\CE_{\rm matter}]=N_f k \left[\text{sign}(v+m_r)-\text{sign}(-v+m_r)\right] =\begin{cases}
0&  v < \vert m_r \vert \\
2N_f k  & \vert m_r \vert < v \\
\end{cases}
\end{split}\end{align}
in accordance with  \cite{Gauntlett:1995fu,Manton:1993aa,Moore:2014gua}.

We now describe how the hyperholomorphic connection on $\CE_{\rm matter}$ arises in 
the supersymmetric quantum mechanics of the collective coordinates.  We choose local coordinates 
$z^m$ on a patch $\CU \subset \froM$, $m= 1, \dots, \dim_{\IR} \froM$,  and a trivialization of $\CE_{\rm matter}$ over that 
patch defined by  a basis $\lambda_s$ of zeromodes of $L$. We can denote these as $\lambda_s(x;z)$ 
where $x\in \IR^3$ and the index $s$ runs from $1$ to the real dimension of $\CE_{\rm matter}$.
In these coordinates  the connection form can be written as
\begin{align}\label{eq:Amnab}
\begin{split}
A_{m,ss'}(z)=\int_{\IR^3} d^3x \text{ }\left\langle \lambda_s(x;z), \left(\frac{\partial}{\partial z^m}+\CR(\epsilon_m)(x;z)\right)\lambda_{s'}(x;z)\right\rangle
\end{split}\end{align}
 where $\langle$ , $\rangle\to \mathbb{R}$ is the canonical Hermitian form on the fibers of $S\otimes E_\CR\to \IR^3$  and $\epsilon_m$ are the components of the universal connection 
 as defined under \eqref{eq:deltamA}. 
 \footnote{ Here we are using the notation $\CR$   both for the 
carrier space of the representation as well as for the homomorphism from $ G $ to 
the general linear transformations of that carrier space.}

\subsection{String Theory Interpretation}\label{subsec:Strings}
\begin{figure}[h]
\begin{center}
\includegraphics[scale=0.65,trim={2cm 12cm 3cm 4cm},clip]{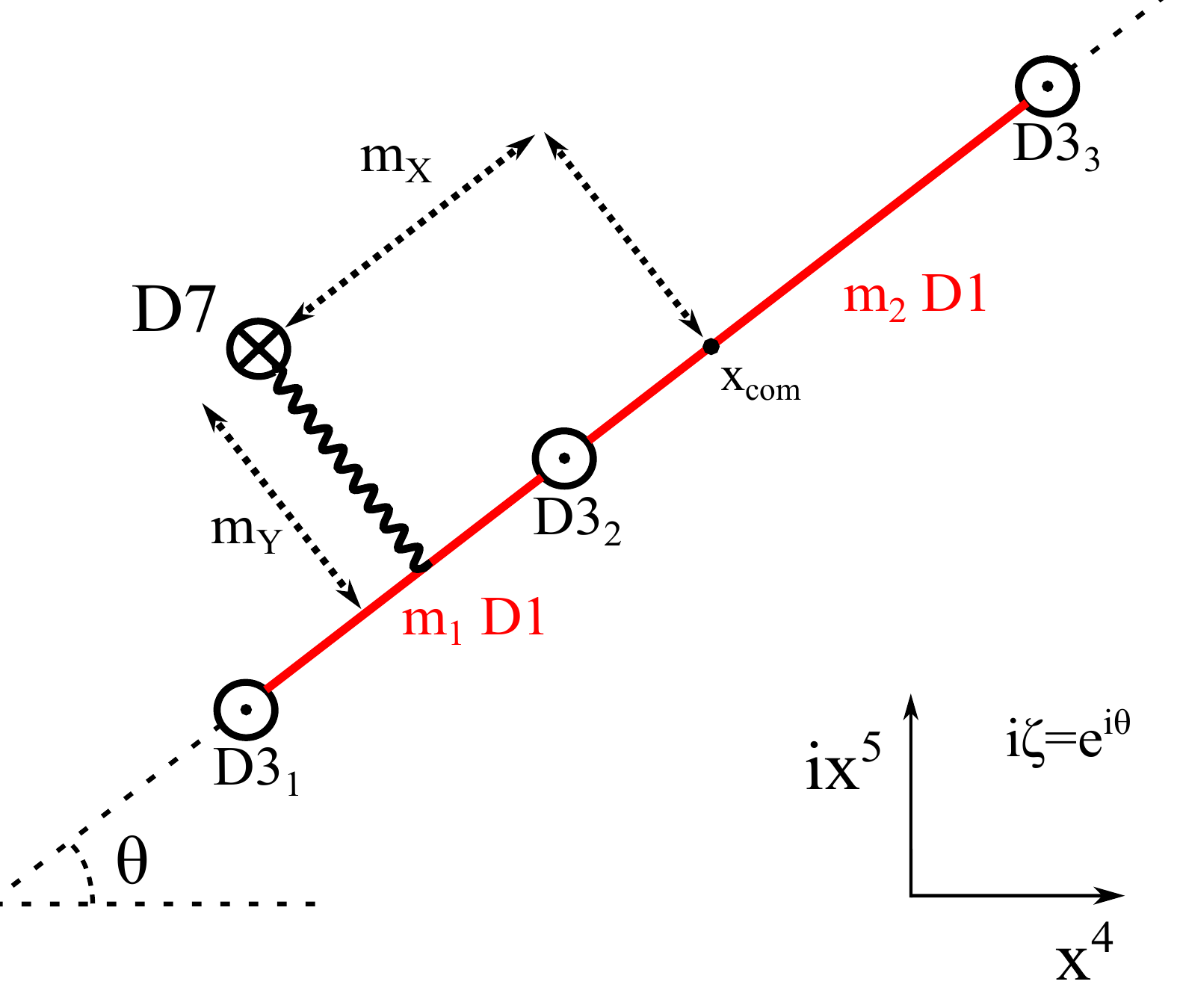}
\caption{This figure illustrates the D-brane configuration corresponding to an SU(3) gauge theory with BPS monopoles of total magnetic charge $\gamma_m=m_1 H_{\alpha_1}+m_2H_{\alpha_2}$ coupled to a single hypermultiplet in the fundamental representation with mass $\zeta^{-1}m=m_y+i m_x$. Here have identified  
 $\IR_{x^4} \oplus\IR_{x^5} \cong\IC$ under $(x^4,x^5)\mapsto x^4+ix^5$. This figure shows that supersymmetric configurations of D3-branes must be collinear in $\IC$, at an angle  
 described by a phase $\zeta\in U(1)$,  and that $m_Y,m_X$ describe the displacement of the D7-brane from the center of mass of the D3-branes. Note that this is consistent with the definition of $\zeta^{-1}=-\frac{Z^{cl}}{|Z^{cl}|}$ as in (B.4). }
\end{center}
\end{figure}
Many aspects of magnetic monopole theory have beautiful geometrical interpretations in terms 
of D-brane configurations. Following the work of \cite{Diaconescu:1996rk} we know that a system of monopoles in a d=4 N=2 SU(N) gauge theory can be realized by a system of N+1 parallel D3-branes  with D1-branes stretched between them. We can then couple this theory to hypermultiplets by introducing D7-branes. This picture will give a geometric interpretation of the phase $\zeta$ and of the hypermultiplet mass parameters  $m_x$ and $m_y$. 

Consider a   system of N+1 D3-branes where the $i^{th}$ brane (the D3$_i$-brane) is localized at $x^6=x^7=x^8=x^9=0$ and a fixed value $x^4+ ix^5=v_i$ such that $v_i/v_j\in \IR$ for all $v_j\neq0$. These values, $v_i$, encode the expectation value of the adjoint valued scalar field in the d=4 N=2 SU(N) vectormultiplet. This tells us that the D3-branes form a straight line in the $x^4+ix^5$ plane whose angle is encoded by $\zeta$. The D1-branes are localized at points $(x^1,x^2,x^3)=\vec{x}_j$ with $x^6=x^7=x^8=x^9=0$. The D1-branes are described in the effective theory of the D3-branes as magnetic 
monopoles located at $\vec x_j$. More accurately, they are D3-tubes running between the D3$_i$-branes.

We can now introduce hypermultiplets by adding D7-branes localized at 
definite values of $x^4+i x^5$, denoted by $m^{(v)}$. The strings stretching between the D7-branes and the D3-tubes support hypermultiplet fields. As such the length of these strings determine the mass of the lowest energy excitations. However, when there are multiple monopoles stretching between the nearest pair of D3-branes there are correspondingly many lowest energy excitations. These count the number of hypermultiplet zero modes and hence the index of the Dirac operator $L$ coupled to the mass $m_X=$Re$[\zeta^{-1}m^{(v)}]=\I m_x$. The jumping of the index of the Dirac operator due to this coupling to $m_X$ implies that $m_X$ should be thought of as the displacement of the D7-brane along the line of D3-branes relative to the center of mass. Similarly  $m_Y=$Im$[\zeta^{-1}m^{(v)}]=-\I m_y$ should be identified with the orthogonal distance of the D7-brane to the line of D3-branes. See Figure 1. 
%
%

\section{Application 1: Generalized Sen Conjecture}\label{sec:GenSen}

The discussion here is almost identical to section 4.1 of \cite{Moore:2015qyu} 
so we will be extremely brief. Upon choosing a complex structure for $\CM$ or $\froM$ 
the bundles $\CE_{\rm line}$ and $\CE_{\rm matter}$ become holomorphic 
bundles with holomorphically flat   connections, and hence the same holds 
for the bundle $\CE$ defined in \eqref{eq:TotalCE}. The wavefunction describing the 
BPS state is an $L^2$ section of $\Lambda^*(T^{0,1}\froM) \otimes \CE $, and a suitable combination of two 
collective coordinate supersymmetries is the Dolbeault operator 
\be
\bar \p^{\CY} = \bar \p_{\CE}  + \text{G}(\CY)^{0,1}\wedge \quad , 
\ee
which squares to zero. The $SU(2)_R$ symmetry does not act on $\CE$ 
because the hypermultiplet fermions are $SU(2)_R$ singlets and the line 
defect preserves $SU(2)_R$ symmetry. Hence the $SU(2)_R$ symmetry 
acts as a holomorphic Lefshetz $\fs\fl(2)$, exactly as in \cite{Moore:2015qyu}, equation (4.3). 
From the no-exotics theorem, conjectured in \cite{Gaiotto:2010be} and 
partially proven in \cite{Chuang:2013wt,CordovaDumitrescu,DelZotto:2014bga}, 
we learn that the $L^2$ cohomology of $\bar \p^{\CY}$ is primitive and 
concentrated in the middle degree.  

It is well-known that the (singular) monopole moduli space can be
formulated, as a complex manifold, as a space of (meromorphic)
maps from $\IC \IP^1$ to $G_{\IC}/B$ where $G_{\IC}$ is the complexification
of $G$ and $B$ is a Borel subgroup. It is natural to formulate the
line and matter holomorphic bundles in these terms, especially when
stating the  generalized Sen conjecture. We expect to explain this
on another occasion. 

As an interesting special case we consider $G=SU(2)$ with a hypermultiplet in the adjoint representation 
of mass $m$, and we take  the $m\to 0$ limit. We take $m_y=0$ and for sufficiently 
small $m_x$ the bundle does not jump.  In this case we can identify $\CE_{\rm matter}$ with the holomorphic tangent bundle. It then follows that BPS states can be described by the self-dual harmonic forms on 
moduli space and in this way we recover the renowned prediction of 
Ashoke Sen based on $S$-duality \cite{Sen:1994yi}. We have thus made good on 
the promise at the end of Section 4.1 of \cite{Moore:2015qyu}. 

\section{Application 2: Explicit Formulae For $L^2$-Indices On Some Monopole Moduli Spaces}\label{sec:Explicit}

It was shown in \cite{Gaiotto:2010be} 
that if   a half-supersymmetric line defect is wrapped on a thermal circle when the 
theory is put on $\IR^3 \times S^1$ then the vev can be expanded as 
\be
\langle L \rangle = \sum_{\gamma \in \Gamma_L } \fro(L;\gamma;u)  \CY_{\gamma}
\ee
where $\Gamma_L$ is a torsor for the IR charge lattice $\Gamma$ and $\CY_{\gamma}$ 
are (locally defined) ``Darboux functions'' of the vacuum of the theory on $\IR^3 \times S^1$.  (That is, they are locally defined functions on  Seiberg-Witten moduli 
space - the total space of the abelian variety fibration over the Coulomb branch given 
by special geometry.) The ``Darboux functions'' obey the  twisted group law: 
\be
 \CY_{\gamma_1}  \CY_{\gamma_2} = (-1)^{\llangle \gamma_1, \gamma_2 \rrangle} 
 \CY_{\gamma_1 + \gamma_2}
\ee
using the electric-magnetic inner product on $\Gamma$. In addition, one can   
``retwist'' by $(-1)^{2I_3}$, where $I_3$ is a generator of $SU(2)_R$ to obtain: 
\be
\langle L \rangle' = \sum_{\gamma \in \Gamma_L } \fro(L;\gamma;u)' \tilde\CY_{\gamma}
\ee
with \emph{retwisted Darboux functions} 
\be
\tilde\CY_{\gamma_1} \tilde\CY_{\gamma_2} = 
\tilde\CY_{\gamma_1 + \gamma_2}.
\ee
Given the no-exotics property, we interpret $\fro(L;\gamma;u)' $ as the dimension of the 
space of framed BPS states and $\fro(L;\gamma;u)$ as the trace of $(-1)^{2J_3}$ over this 
space. 

As an application   of our general result for the semiclassical interpretation of $\fro(L;\gamma;u)$ 
when general line defects are included we specialize to the $G=SU(2)$ theory with $N_f=0$ 
and with $P=0$ and $Q= \frac{\kappa}{2} \alpha$ where $\kappa$ is a positive integer. 
Recall that $\Lambda_{wt} = \frac{\alpha}{2} \IZ$ and $\Lambda_{mw} =  \frac{H_{\alpha}}{2} \IZ$ 
so that, in the $SU(2)$ theory, 
\be
\Gamma_L = \frac{\kappa}{2} \alpha + \Gamma = \Gamma = 2\Lambda_{mw} \oplus \Lambda_{wt}.
\ee
Then the supersymmetric Wilson line is 
\be
W_{\kappa/2}:=
L_{\zeta}[0,\frac{\kappa}{2} \alpha ] = \Tr_{\kappa+1}\Biggl( P\exp\oint \left[ A - Y \right] \Biggr)
\ee
where here and below the subscript on the trace indicates the dimension of an irreducible 
representation, and the adjoint scalar $Y$ is defined in equation \eqref{eq:XY-Field-Def} below. 

Since  this is a theory of class S the quantum vev $\langle L \rangle$ is 
given by the classical holonomy of a complex flat connection on the underlying 
UV curve $C$. (See \cite{Gaiotto:2010be}, section  7.4.  The  flat connection on $C$   encodes the vacuum 
on $\IR^3 \times S^1$ and the holonomy is taken around a closed loop on $C$ encoding the 
line defect $L$.)  Now, for any group element $h$  in $SL(2,\IC)$ 
 the trace in the irreducible representation 
of dimension $\kappa+1$ is related to that in the fundamental representation 
by  
\be
\Tr_{\kappa+1}(h) =  U_{\kappa} \left(\half \Tr_{2}(h) \right) 
\ee
where   $U_n$ is the 
Chebyshev polynomial of the second kind, satisfying: 
\be
U_n(\cos\theta) = \frac{\sin[ (n+1)\theta]}{\sin\theta}.
\ee
Therefore: 
\be\label{eq:GenSU2-Wilson}
\langle W_{\kappa/2} \rangle  =U_{\kappa}\left( 
\half \langle W_{1/2} \rangle \right).
\ee
Now equation (10.33) of \cite{Gaiotto:2010be} gives  an explicit expression:
\be
\langle W_{1/2} \rangle  =  \CY_{\half \alpha} +  \CY_{-\half \alpha} +  \CY_{H_{\alpha}+\half (2c+1)\alpha }
\ee
where $c\in \IZ$ labels chambers separated by the ``BPS walls'' where 
framed BPS states jump, and we work in a 
semiclassical domain where $\CY_{H_\alpha}$ is exponentially small as the coupling 
goes to zero.  (The dependence on chamber comes about from the value of $Y_\infty$, 
and from the lift of $\zeta$ to the universal cover of $U(1)$.)
There is a parallel expression for 
$\langle W_{\kappa/2} \rangle'$ with $\CY_{\gamma} \rightarrow \tilde \CY_{\gamma}$. 

Since $U_{\kappa}(x)$ is a polymomial in $x$ it is, in principle, straightforward to expand 
\eqref{eq:GenSU2-Wilson}  to compute $\langle W_{\kappa/2}\rangle$ as an expansion in $\CY_{\gamma}$
and thereby obtain $\fro(W_{\kappa/2};\gamma; u)$. On the other hand, given the results of the present paper,
 for $u$ in the weak-coupling chambers we can interpret the 
framed degeneracies  as characters of an $L^2$- kernel of a Dirac-like operator on $\CM(\gamma_m;\CX)$. 
We spell out the identification in detail as follows: 

The Dirac operator acts on spinors on $\CM(\gamma_m ; \CX)$ 
(with    $\gamma_m = m H_{\alpha} $)
coupled to the hyperholomorphic bundle $\CE_{\rm line }$. The bundle 
$\CE_{\rm line }$ in this case  is just the universal bundle 
in the $(\kappa+1)$-dimensional irreducible representation of $SU(2)$, restricted to 
$\{ \vec x \} \times \CM(\gamma_m;\CX)$ (for any fixed $\vec x$).  
If $\CY=0$  we use the standard Dirac operator;  in general we add Clifford multiplication by 
$\slashed{\rmG}(\CY)$ as in equation \eqref{eq:DiracOperator}. 
\footnote{N.B. Here $\CY$ is the  vev of a Higgs field, as in \eqref{eq:XYdef}  and should not be confused 
with a Darboux function $\CY_{\gamma}$ !}

Now suppose that $\gamma = m H_\alpha + \frac{n_e}{2} \alpha$. Then BPS states of this 
charge are located in the   $\frac{n_e}{2} \alpha$-isotypical component  of the kernel of the Dirac operator. 
This means that 
on   $\CM (\gamma_m ; \CX)$ 
 the Lie derivative of the spinor under the hyperholomorphic vector field
$\text{G}(H_\alpha)$    acts as  
\be 
\CL_{\text{G}(H_\alpha) } \Psi =  -\I n_e \Psi .
\ee
Now, $\fro(W_{\kappa/2} ; \gamma;u)$ is the trace of $(-1)^{2J_3}$  in the 
$\frac{n_e}{2} \alpha$-isotypical component while the retwisted degeneracy 
$\fro(W_{\kappa/2} ; \gamma;u)'$  is   just the dimension of that   component.

Expanding the Chebyshev polynomial in power series and rearranging a little we 
find 
\be
\langle W_{\kappa/2} \rangle = 
\sum_{m=0}^{\kappa} \sum_{
n_e } 
\fro(W_{\kappa/2}, \gamma = m H_\alpha + \frac{n_e}{2}\alpha ;u)
  \CY_{m H_\alpha + \frac{n_e}{2}\alpha} 
\ee
where the sum over $n_e$ only includes integers with $n_e = \kappa \mod 2$ in the 
range
\be
2m - \kappa \leq (n_e - 2mc) \leq \kappa 
\ee
and 
\be\label{eq:BigSum}
\fro(W_{\kappa/2}, \gamma = m H_\alpha + \frac{n_e}{2}\alpha  ;u)
=(-1)^{\kappa (m-\kappa)}  \sum_{\ell \geq0 } (-1)^{\ell} 
\binom{\kappa-\ell}{\ell} \binom{\kappa - 2\ell}{m} \binom{\kappa-2\ell - m}{ N_c-\ell}
\ee
with 
\be
N_c  := \frac{ \kappa - (n_e  - 2mc)}{2}   = \frac{ \kappa - n_e  }{2}  + mc.
\ee
Note that $N_c\in \IZ$. 
In \eqref{eq:BigSum}  the summands vanish unless 
\be
0 \leq \ell \leq {\rm Min}[N_{c} , \kappa - N_{c+1} ] .
\ee

We can similarly expand $U_{\kappa}\left( \half \langle W_{\kappa/2}\rangle' \right)$ 
using the retwisted   Darboux functions $\tilde \CY_\gamma$ 
and we find:  
\be
\fro(W_{\kappa/2}, \gamma = m H_\alpha + \frac{n_e}{2}\alpha  ;u) '
= (-1)^{m(\kappa  -m)} \fro_\CY(W_{\kappa/2}, \gamma = m H_\alpha + \frac{n_e}{2}\alpha  ;u)
\ee
This has the interesting consequence that for fixed charge $\gamma$ all the BPS states are 
either fermionic or bosonic, the parity being determined by the parity of $m(\kappa - m)$. 

In fact, it is possible to simplify \eqref{eq:BigSum} by recognizing it as a special value of 
a hypergeometric series leading to the elegant result for the framed BPS degeneracy in 
the chamber labeled by $c\in \IZ$: 
\be\label{eq:BigSum2}
\fro(W_{\kappa/2}, \gamma = m H_\alpha + \frac{n_e}{2}\alpha  ;u)' 
= \binom{ N_{c+1}  }{m} \binom{  \kappa - N_{c }  }{m} .
\ee
For fixed $m$ and $c$ this is a polynomial in $\kappa$ of order $2m$, suggestive of an 
index theorem on the (noncompact) monopole moduli space of dimension $4m$. 

\subsection{Marginally Bound States: Remark On A Paper Of Tong And Wong} 

Framed BPS states in N=2 gauge theory in the presence of a Wilson line 
have been previously studied by Tong and Wong in \cite{Tong:2014yla}. These authors also point 
out that inclusion of Wilson lines leads to a modification of the relevant 
Dirac operator by coupling to a bundle with connection.  However, the paper 
\cite{Tong:2014yla} raised a puzzle because 
there is a slight discrepancy between their equation $(4.10)$ and the results 
of \cite{Gaiotto:2010be}. In this subsection we explain that the source of the 
discrepancy can be traced to how one handles marginally bound states. 

 In our notation, equation 
$(4.10)$ of  \cite{Tong:2014yla} can be written as  
\be\label{eq:TW2}
\fro(W_{\kappa/2}, \gamma = H_\alpha + \frac{n_e}{2} \alpha )'=\half \sum_{\substack{|s|\leq \kappa\\s\in 2\IZ+\kappa}}s \text{ sign}(s-n_e-n_e\epsilon)
\ee
where $n_e\in 2\IZ+\kappa$ and $\epsilon>0$ is an infinitesimal regularizing parameter.
One way to obtain an analogous result from our expressions is to write 
\be
\langle W_{1/2} \rangle' = \xi + \xi^{-1}
\ee
with 
\be
\xi = \half \left( \tilde \CY_{\half \alpha} +\tilde  \CY_{-\half \alpha} +\tilde  \CY_{H_\alpha+\half\alpha } 
+ \sqrt{ ( \tilde  \CY_{\half \alpha} -\tilde  \CY_{-\half \alpha} )^2+ 2(\tilde \CY_{\half \alpha} + \tilde \CY_{-\half \alpha} )\tilde \CY_{H_\alpha +\half \alpha}  + \tilde \CY_{2H_\alpha+\alpha } } \right)
\ee
and then expand 
\be
\langle W_{\kappa/2} \rangle = \sum_{\substack{s= - \kappa\\s\in 2\IZ+\kappa}}^{\kappa} \xi^s .
\ee
%
This gives, for example,  
\be
\begin{split}\label{eq:BM2}
\fro(W_{\kappa/2}, \gamma=   H_\alpha + \frac{n_e}{2}\alpha  ;u)'& = 
 \sum_{\substack{s \geq |n_e| \\s\in 2\IZ+ \kappa}}^\kappa s \Theta(s-1 -\vert n_e -1 \vert) \\
\end{split}
\ee
where $\Theta(x) \in \{0,1\}$ is the Heaviside step function and $\Theta(0)=1$. We can then 
rewrite this equation in a form analogous to \eqref{eq:TW2}:  
\be\label{eq:BM3}
\fro(W_{\kappa/2}, \gamma = H_\alpha + \frac{n_e}{2} \alpha )'=\half \sum_{\substack{|s|\leq \kappa\\s\in 2\IZ+\kappa}}s \text{ sign}(s-n_e\pm \epsilon)
\ee
where the choice of $\pm\epsilon$ is equivalent to choosing a chamber in which:
\be
\langle W_\half\rangle=\CY_{\half \alpha}\oplus \CY_{-\half \alpha}+\CY_{H_\alpha \pm \half\alpha}.
\ee

 In  \cite{Tong:2014yla} the framed BPS states with magnetic charge $m=1$ 
are counted via an index theorem, but the relevant Dirac operator is not Fredholm. 
The Dirac operator is evaluated for the theory at a wall of marginal stability. Physically, 
as explained in \cite{Tong:2014yla}, 
one must worry about whether or not to include marginally bound states. 
The expression \eqref{eq:TW2} makes use of one perturbation to a Fredholm operator.  
However the result  conflicts with the general computation of \eqref{eq:BM2} and \eqref{eq:BigSum2}
and hence the counting of boundstates used in  \cite{Tong:2014yla} differs from that 
used in   \cite{Gaiotto:2010be}.  Another way to perturb to a Fredholm operator is to 
turn on   a small generic $\CY$. This changes the perturbation, $n_e \to n_e + \epsilon n_e$,
used in \eqref{eq:TW2}, to the perturbation $n_e \to n_e \mp \epsilon$, used in 
\eqref{eq:BM3}. The latter perturbation brings the index into line with 
 the general results of \cite{Gaiotto:2010be}. 

\appendix

\section{Review: The Universal Bundle And The Universal Connection}\label{app:Universal}

In this section we review the universal bundle   of Atiyah and Singer \cite{Atiyah:1984tf}. 
(An expository account can be found in many places, among them \cite{Cordes:1994fc} sec. 8.8.) 

 Let $G$ be a compact, semisimple Lie group with a trivial center and let $\pi: \fP\to M$ be a principal $G$ bundle and let $\mathcal{G}=\{\Phi:\fP\to \fP \vert \pi\circ \Phi=\pi \}$ be the group of gauge transformations (bundle automorphisms). Let $\CA$ be the space of suitably 
smooth connections on $\fP$. The group $\mathcal{G}$ acts on $\fP\times\mathcal{A}$ by:
\begin{align}\begin{split}
  (p,A)\cdot g =(p\cdot g ,g^{-1} A g+g^{-1}dg)\hspace{10mm} \hspace{5mm}(p,A)\in \fP\times \mathcal{A}
\end{split}\end{align}
(Where $p\cdot g$ means the right-action on the principal $G$ bundle by the value of the gauge 
transformation $g$ at the point $\pi(p)\in M$.) 
We would like to form the $G \times \CG$ bundle $\pi: \fP \times \CA \to M \times \CA/\CG$ but 
because the group action can fail to be free we replace $\CA$ by a space $\CA^*$ 
whose \emph{raison d'etre} is to have a free action. 
 \footnote{One choice of $\CA^*$ is simply the subspace of $\CA$ on which $\CG$ acts freely. 
 Another maneuver replaces the group $\CG$ of gauge transformations by the subgroup 
 fixing a point in $\fP$.  Yet another  
 choice is to consider the space of \underline{framed} bundles with connection.  A framing is a choice of basepoint $x_0\in M$ together with a $G$-equivariant map $\varphi:G\to \fP_{x_0}$. Denote 
 the space of these equivariant maps by $\Hom(G,\fP_{x_0})$. Then we can take  
 $\CA^*:= \CA \times \Hom(G, \fP_{x_0})$. (In this case one must modify some of the formulae 
 for the tangent space below - in a straightforward way.) Similar considerations show that if  
 we were to include groups with a non-trivial center we would need to restrict $\fP$ to be a principal $G_0$ bundle where $G_0=G/Z(G)$. }

Note that we have the diagram of projections:  
\begin{align}\begin{split}
\xymatrixcolsep{0.5pc}\xymatrixrowsep{10mm}
\xymatrix{
&\fP\times\mathcal{A}^*\ar[dr]^G\ar[dl]_{\CG}&\\
\CQ=\fP\times \mathcal{A}^*/\mathcal{G}\ar[dr]_G&&M\times \mathcal{A}^*\ar[dl]^{\CG}\\
&M\times\mathcal{A}^*/\mathcal{G}&
}
\end{split}\end{align}
The principal $G$-bundle $\pi: \CQ\to M\times\mathcal{A}^*/\mathcal{G}$ was referred to by Atiyah and Singer as the \emph{universal bundle}, (and indeed it enjoys a universal property). Another useful bundle is   the principal $G\times \CG$-bundle with total space $\fQ := \fP \times \CA^* $ and projection 
 $\pi: \fQ \to M \times \CA^*/\CG$.  
 
The bundle $\mathfrak{Q}$ has a natural connection which we will refer to (by a slight abuse of 
terminology) as the 
\emph{universal connection}. To define it, note that, given a metric on $M$ and a Killing metric $\Tr(...)$
on $\fg$ there is a natural metric on $\CA$ defined by 
\be\label{eq:CA-metric}
(\tau_1, \tau_2) = \int_M \vol  \Tr(\tau_1 * \tau_2 ) 
\ee
where $\tau_i \in T_A \CA \cong \Omega^1(M;{\rm ad}(\fP))$, with $i=1,2$. 
Now, a connection can be defined by specifying the horizontal subspaces of $\fQ$ in 
the tangent space $T_{p,A} \fQ \cong T_p \fP  \oplus T_A \CA$ orthogonal to the 
subspace of vertical vectors $\cong \fg \oplus \Lie(\CG)$. The horizontal subspace is 
defined by 
\be
H_{p,A} := H_p(A) \oplus \Lie(\CG)^\perp
\ee
where $H_p(A) \subset T_p \fP$ is the horizontal subspace determined by the connection $A$ 
and $\Lie(\CG)^\perp$ is the orthogonal complement to the infinitesimal gauge transformations 
in the metric \eqref{eq:CA-metric}. 

Note that a very similar construction also gives a connection on $\pi: \CA^* \to \CA^*/\CG$, 
namely, the horizontal subspaces are the orthogonal subspaces to the gauge orbits in the metric \eqref{eq:CA-metric}. 
By an even more abusive use of terminology we will \underline{also} refer to this connection as the 
``universal connection.'' 
It will be useful to be more explicit about this connection: If $\tau = \frac{d}{dt} A(t)$ is a tangent 
vector at $A \in \CA^*$ then since the   vertical vector fields in $T_A \CA^*$ are associated to  $\epsilon: M \to \Omega^0(M; {\rm ad}(\fP))$ 
and  given by $\tau_{\epsilon}:=   - D_A \epsilon$ the horizontal projection of $\tau$ is 
\be\label{eq:Hproj}
H(\tau) = \tau - D_A \tilde\epsilon 
\ee
where $\tilde \epsilon$ is the unique solution to 
 $D_A * (\tau - D_A \tilde \epsilon) =0$ vanishing at the framing point $x_0$.

In the application to magnetic monopoles we take $M=\IR^3$ with Euclidean metric 
and choose $x_0$ to be a point at infinity (chosen along a particular direction). 
The ``connections'' in $\CA$ are actually translationally invariant connections $\hat A$ on $M \times \IR$, 
and we interpret 
\be\label{eq:Ahat}
\hat A = A_i dx^i + X dx^4
\ee
where $X$ is the Higgs field valued in $\fg$. 
One often pulls back the bundle to $\CM \subset \CA^*/\CG$. In this context, if 
$\hat A(z^m)$ is a family of gauge-inequivalent solutions to the Bogomolnyi equations 
parametrized by an open set of $\CM$ with local coordinates $z^m$, $m =1, \dots, \dim_{\IR} \CM$ 
then 
\be\label{eq:deltamA}
\tau_m= 
\frac{\p \hat A}{\p z^m} 
\ee
is in general not in the horizontal subspace 
and the compensating gauge transformation $\tilde\epsilon$ defined above 
is denoted $\epsilon_m$, with horizontal projection $H(\tau_m):=\delta_m \hat A$. 
This defines notation used in equation \eqref{eq:Amnab}   above 
and in appendix \ref{sec:SQM} below.  

Finally, in the case of singular monopoles, the connection $\hat A = (A, X)$ must satisfy 
the boundary conditions \eqref{eq:UV-BC}  at the location of each line defect $\vec x_j$. As explained 
carefully in \cite{Kapustin:2005py,Moore:2015szp}
this means there is a reduction of structure group at 
$\{ \vec x_j \} \times \CA^*/\CG$ to $Z(P_j)\subset G$, the centralizer of the 't Hooft charge 
at $\vec x_j$. Note that if $P=0$ then $Z(P)=G$.

\section{Proof Using Supersymmetric Quantum Mechanics}\label{sec:SQM}

In this appendix we provide a few of the details of the formulation of the collective 
coordinate supersymmetric quantum mechanics which is the basis of the above 
formulation of the semiclassical space of BPS states. 

The UV Lagrangian written in d=4 N=1 superspace is  (using standard notation, such 
as in  \cite{Labastida:2005zz}): 
\begin{align}\begin{split}
\mathcal{L}=&\left[-\frac{\I \tau}{4\pi}\int d^2\theta \Tr( W_{\alpha}W^\alpha)+c.c.\right] +\frac{\text{Im }\tau}{4\pi}\int d^4\theta \text{ }  \Phi^\dagger e^{2i V}\Phi  \\
&+\frac{\text{Im }\tau}{4\pi}\left\{\int d^4\theta \left(Q^{\dagger} e^{2i V}Q +\tilde{Q}^{\dagger} e^{-2iV}\tilde{Q}\right)+
\left[ \int d^2 \theta (i \tilde{Q} \Phi Q +  i\tilde{Q} m Q )+c.c.\right] \right\}
\end{split}\end{align}

Here we have assumed $G$ is a simple group and $W_\alpha$ is the chiral superfield 
associated to an $N=1$ vectormultiplet for $G$, while $\Phi$ is a chiral superfield in the adjoint of $G$.
\footnote{Here $\alpha$ is traditional notation for a spinor index and has nothing to do with 
the root $\alpha$ of $\fs\fu(2)$ used elsewhere in this note.}
 Note that the  N=1  superspace formalism implicitly assumes a splitting of the quaternionic representation 
of $G_f \times G$ in the form $\CR = \rho \oplus \rho^\ast$ with  chiral superfields $Q$ and $\tilde{Q}$ 
transforming in representations $\rho$ and $\rho^*$, respectively. Finally, without loss of generality 
we can assume $m \in \ft_f\otimes \IC$.

We next write out the Lagrangian in terms of the components of the superfields. 
The lowest component of $\Phi$ is the scalar $\varphi$ valued in $\fg\otimes \IC$. 
It is convenient to decompose the 
vectormultiplet  scalar fields  and mass parameters into ``real'' and ``imaginary'' parts 
according to 
\be\label{eq:XY-Field-Def}
\zeta^{-1} \varphi = Y + \I X\qquad Y, X \in \fg
\ee
\be
\zeta^{-1} m = m_y + \I m_x  \qquad  m_y, m_x \in \ft_f.
\ee
As noted above,  $\zeta$ will be the phase defining the line defect, or, if there is no line defect, 
it is the phase of the classical limit of the central charge 
\be\label{eq:Zdef}
\zeta^{-1} = - \frac{Z^{cl} }{ \vert Z^{cl} \vert} \qquad  Z^{cl} = \frac{4\pi \I}{g_0^2} ( Y_\infty + \I X_\infty, \gamma_m) 
\ee
where $X_\infty$ and $Y_\infty$ are vacuum expectation values of $X,Y$ at $\vec x\to \infty$. 
That is, we have boundary conditions at infinity: 
\be\label{eq:3}
\begin{split}
X  & = X_\infty - \frac{\gamma_m}{2r} + \cdots \\
F  & = \half \gamma_m \omega + \cdots \\
\end{split}
\ee
and $Y \to Y_\infty + \cdots $ (compatible with the equations of motion) to leading order in a 
large $r$ expansion.  
The vevs $X_\infty, Y_\infty$ are related to the vevs $\CX,\CY$ used elsewhere in this note by:  
\be
\CX=X_\infty\hspace{10mm}\CY=\frac{4\pi}{g_0^2}Y_\infty+\frac{\theta_0}{2\pi}X_\infty . 
\ee
If we use the definitions \eqref{eq:XYdef} then these are only the leading expressions in a 
weak coupling expansion. (It was argued in \cite{Moore:2015szp} that the higher order terms 
in the definition \eqref{eq:XYdef} 
correctly capture perturbative corrections to the collective coordinate dynamics.)

The 't Hooft-Wilson operator $L_\zeta[P_j,Q_j]$, inserted at a point $\vec x_j$ 
modifies the path integral in two ways: 

\begin{enumerate}

\item 
First, it modifies the boundary conditions on the fields over which we integrate. 
We choose a representative of 
$[P_j,Q_j]$ so that $Q_j$ is a dominant weight of $Z(P_j)$ 
and impose boundary conditions near $\vec x_j$: 
\begin{align}\label{eq:UV-BC}
\begin{split}
B^i&=\frac{P_j}{2r_j^2}\hat{r}_j^i+O(r_j^{-3/2})\hspace{10mm}E^i=\frac{g_0^2}{4\pi}\frac{Q_j^\ast}{2 r_j^2}\hat{r}_j^i-\frac{\tilde\theta_0 P_j}{2 r_j^2}\hat{r}_j^i+O(r^{-3/2})\\
X&=-\frac{P_j}{2 r_j}+O(r_j^{-1/2})\hspace{11mm}Y=-\frac{g_0^2}{4\pi}\frac{Q_j^\ast}{2 r_j}+\frac{\tilde\theta_0 P_j}{2 r_j}+O(r_j^{-1/2})
\end{split}\end{align}
where $r_j = \vert \vec x - \vec x_j \vert$,  
 $Q_j\in \Lambda_{wt}$ is the highest weight of a representation $R_j$  of $Z(P_j)$, and  $Q_j^\ast\in \ft$ is the dual of $Q_j$ under the canonical pairing $\langle$ , $\rangle:\ft^\vee \times\ft\to \IR$. 

\item 
Second, we insert a quantum mechanical path integral, representing modes located at the 
position of the line defect.  Let $R_j \cong \IC^{N_j}$ denote the 
irreducible representation with highest weight $Q_j$. We may assume it is a unitary representation 
with the standard Hermitian metric. 
We introduce $N_j$ complex fermions   $w_j \in R_j $ and introduce the action 
\be\label{eq:WilsonAction}
S_{j}=\int d^4x\text{ }\delta^{(3)}(x-x_j)  \left[iw_j^\dagger\left(\p_t+ R_j(A_{0}-Y) -\I 
\alpha_j(t)\right)w_j + \frac{N-2}{2}\alpha_j(t) \right]
\ee
Here and below we use the notation   $R_j(F)$ to indicate that a  $\fg$-valued field $F$ 
is evaluated at $\vec{x}_j$ and then represented in the  $R_j$ representation. 
We are again using the notation for $R_j$ 
in a way similar to that explained in the footnote under equation \eqref{eq:Amnab}. 
In equation \eqref{eq:WilsonAction} note that the pole structure of $A_0$ and $Y$ 
do not always allow them to be defined at the the points $\vec{x}_j$, but their difference will always be well defined. 
Finally,   $\alpha_j(t)$ is a 
Lagrangian multiplier enforcing the constraint that in the Hilbert space we project onto 
the one-particle sector for the number operator 
\be
\half ( w_j^\dagger w_j - w_j w_j^\dagger ) .
\ee

\end{enumerate}

We now introduce collective coordinates. We 
choose a local patch in $\CM$ or $\froM$ with coordinates $z^m$, $m=1\dots, \dim_{\IR}\CM$ or 
$\dim_{\IR} \froM$
 and promote these to time-dependent fields. Then we try to solve the classical equations of motion. 
The solution of the Dirac equation for the vectormultiplet fermions  introduces superpartners
$\chi^m(t)$. When we include the hypermultiplets the   hypermultiplet scalars are set to zero 
(for generic point on the Coulomb branch, and 
certainly in the semiclassical limit where $u\to \infty$) and the solution of the Dirac equation   
for the hypermultiplet fermions introduces real fermionic coordinates $\psi^s(t)$ 
with $s = 1, \dots, \dim_{\IR} \CE_{\rm matter}$. 

The result of a fairly long computation is the collective coordinate Lagrangian: 
\footnote{In the systematic weak-coupling expansion of the action one must also 
include a one-loop correction to the mass from vacuum diagrams. This is not included below.}
\begin{align}\begin{split}
L_{c.c.}&=\frac{4\pi}{g_0^2}\left[\frac{1}{2}g_{mn}\left(\dot{z}^m\dot{z}^n+\I \chi^m \mathcal{D}_t\chi^n-\text{G}(Y_\infty)^m\text{G}(Y_\infty)^n\right)+\frac{\I }{2}\chi^m\chi^n\nabla_m\text{G}(Y_\infty)_n\right]\\
&- \frac{4\pi}{g_0^2}\left[\I  h_{ss'}\psi^s \mathcal{D}_t\psi^{s'}-\I \psi^s (m_{y,ss'} + T_{ss'})\psi^{s'}+\frac{1}{2}F_{mn,ss'}\chi^m\chi^n \psi^s\psi^{s'}\right]\\
&-\frac{4\pi}{g_0^2}(\gamma_m,X_\infty)+\frac{\theta_0}{2\pi}(\gamma_m,Y_\infty)+\frac{\theta_0}{2\pi}\Big(g_{mn}(\dot{z}^m-\text{G}(X_\infty)^m)\text{G}(Y_\infty)^n-\I \chi^m\chi^n\nabla_m\text{G}(X_\infty)_n\Big)\\
& +\I \sum_{j}w^\dagger_j (\p_t + R_j(\epsilon_{Y_\infty})-R_j(\epsilon_m) \dot z^m +  \frac{\I}{2}R_j(\phi_{mn})\chi^m\chi^n
-\I \alpha_j(t) )w_j + \frac{N-2}{2}\alpha_j(t)
\end{split}\end{align}
Here $\epsilon_m$ is the compensating gauge transformation used in defining the 
universal connection, as defined under \eqref{eq:deltamA}. The corresponding 
curvature of the universal connection is 
\be\label{eq:phimn}
\phi_{mn} = [D_m, D_n]
\ee
where $D_m = \frac{\p}{\p z^m} + [ \epsilon_m, \cdot]$.
Similarly, for any element $H\in \ft$ we define $\epsilon_H$ to be the 
solution of $\hat D^2 \epsilon_H=0$ with boundary condition 
that $\epsilon_H \to H$ at infinity.
In addition we have: 
\begin{gather*}
g_{mn}=\frac{1}{2\pi}\int_{\mathbb{R}^3} d^3x\text{ Tr}\left\{\delta_m\hat{A}^a \delta_n\hat{A}_a \right\}\hspace{10mm}
\Gamma_{mnp}=\frac{1}{2\pi}\int_{\mathbb{R}^3}d^3x\text{ Tr}\left\{\delta_m\hat{A}^a  D_p\delta_n\hat{A}_a\right\}\\
\mathcal{D}_t\chi^n=\dot{\chi}^n+\Gamma^n_{mp}\dot{z}^m\chi^p\hspace{10mm}
\mathcal{D}_t\psi^s=\dot{\psi}^s+\dot z^m (A_m )^s_{\text{ }s'}\psi^{s'}\\
m_{y,ss'}=\frac{1}{2\pi}\int d^3x\text{ }\overline\lambda_s m_y \lambda_{s'} \hspace{23mm}A_{m,ss'} =\frac{1}{2\pi}\int d^3x\text{ }\overline\lambda_s (\partial_m+\CR(\epsilon_m))\lambda_{s'} \\
T_{ss'} =\frac{1}{2\pi}\int d^3x\text{ }\overline\lambda_s \CR(\epsilon_{Y_\infty})\lambda_{s'} \hspace{10mm} F_{mn,ss'}= \partial_{ m} A_{n,ss'}- \partial_{n}A_{m,ss'}
 +( A_{m }A_{n }-A_{n }A_{m })_{ss'}
\end{gather*}
%
%

One can check that the action is invariant under the   $N=4$   supersymmetry transformations:
\begin{align}\begin{split}\label{eq:Supertmns}
\delta_\nu z^m&=-\I \nu_a (\tilde{\mathbb{J}}^a)^m_{\text{ }n}\chi^n\\
\delta_\nu \chi^m&=(\mathbb{J}^a)^m_{\text{ }n}(\dot{z}^n-\text{G}(Y_\infty)^n)\nu_a-\I \nu_a\chi^k\chi^n(\mathbb{J}^a)_k^\ell\Gamma_{\ell n}^m\\
\delta_\nu\psi^s&=-A_{m,s'}^s\delta_\nu z^m\psi^{s'}\\
\delta_\nu w_j   & =   \delta_\nu z^m R_j(\epsilon_m) w_j \\
\delta_\nu \alpha_j(t) & = 0 
\end{split}\end{align}
where
\begin{align}\begin{split}
\mathbb{J}^a=(\mathbb{J}^r,\textbf{1})\hspace{10mm}\tilde{\mathbb{J}}^a=(-\mathbb{J}^r,\textbf{1})
\end{split}\end{align}
for an index $a=1,2,3,4$ and where 
  $\mathbb{J}^r$, $r=1,2,3$ are three covariantly constant complex structures on $\froM$ (or $\CM$)
satisfying the quaternion relations. 
Note that the number operator for the $w_j$ fermions is invariant under supersymmetry 
transformations so we may restrict to the one-particle sector without breaking supersymmetry. 
The check that the 
action is indeed   invariant under  \eqref{eq:Supertmns}  makes use of the 
property that the   connections on $\CE_{\rm matter}$ and $\CE_{\rm line}$ are   hyperholomorphic. 
Because the collective coordinate Lagrangian \underline{must} have N=4 supersymmetry 
this can be regarded as a proof that these connections are indeed hyperholomorphic.

Upon quantization we find   the supercharge operators:
\begin{align}\begin{split}
\hat{Q}^a&=-\frac{\I g_0}{2\sqrt{2\pi}}\gamma^m(\tilde{\IJ}^a)_m^{\text{ }n}\times\\
&\left(\partial_m+\frac{1}{4}\omega_{m,pq}\gamma^{pq}+\frac{1}{2} \Omega_{m,ss'}\theta^{ss'} -\sum_{j} w_j^\dagger R_j(\epsilon_m^{(j)})w_j-\I \text{G}(\CY_\infty)_m\right)
\end{split}\end{align}
where $\omega_{m,pq}$ is the spin connection for the \hk\  metric on the monopole moduli space 
 $\Omega_{m,ss'}$ is the hyperholomorphic connection on $\CE_{\rm matter}$ and $\theta^s$ 
 are the gamma matrices acting on ${\rm Spin}(\CE_{\rm matter})$ so that 
 $\theta^{ss'} := \theta^s \theta^{s'}$ for $s\not=s'$. 
 \footnote{These come from the quantization of the hypermultiplet fermions 
 $\psi^s = (\frac{g_0^2}{4\pi})^{1/2} \theta^s$. So, in the Hamiltonian formulation 
 $\{ \theta^s, \theta^{s^\prime} \} = 2\delta^{ss^\prime} $ . }

One can check - as expected - that these operators  satisfy the N=4 SQM algebra:
\be
\{\hat{Q}^a,\hat{Q}^b\}=2\delta^{ab}(\hat{H}-\text{Re}(\zeta^{-1}\hat{Z}))
\ee
The central charge satisfies:
\be
\text{Re}(\zeta^{-1}\hat{Z})=M_\gamma^{1-lp}  
\ee
where
\be
\hat Z = (\gamma_m, a_D) + \hat \gamma_e \cdot a + \hat \gamma_f \cdot m
\ee
and $\hat \gamma_e$ and $\hat \gamma_f$ are operators in the quantum mechanics. 
The operator $\hat \gamma_e$ is defined by the generators of the global gauge transformations in $T$ 
(see \cite{Moore:2015szp} for the detailed expressions) while 
\be\label{eq:FlavorCharge}
\hat  \gamma_f\cdot m_y = i\theta^s m_{y,ss'} \theta^{s'} .
\ee
When representing the Clifford algebra $\theta^s$ we must choose 
a proper normal-ordering constant, and this must be determined by 
physical considerations. For example, in the string theory interpretation 
of section \ref{subsec:Strings} it should represent the energy from the 
tension of fundamental strings stretched between the D7 and D3 branes. 

It follows from the supersymmetry algebra that a wavefunction is in the kernel of   $\hat Q^a$ 
either for all operators $a=1,2,3,4$ of for none of them.  
Therefore, it suffices to focus on the Dirac operator proportional to $\hat Q^4$: 
\be
\I \gamma^m\left(\partial_m+\frac{1}{4}\omega_{m,pq}\gamma^{pq}+\frac{1}{2} \Omega_{m,ss'}\theta^{ss'} -\sum_{j} w_j^\dagger R_j(\epsilon_m)w_j- \I  \text{G}(\CY)_m\right)
\ee
The BPS states with magnetic charge $\gamma_m$ are the $L^2$ wavefunctions on $\CM$ or $\froM$ 
in the kernel of $\hat Q^4$.   The subspaces with definite electric and flavor charge 
are the isotypical subspaces of the $T_f \times T$ action on the kernel -  equivalently - the eigenspaces of 
$\hat \gamma_e$ and $\hat \gamma_f$.

\end{document}